\documentclass[prl,showpacs,twocolumn]{revtex4}
\usepackage{bm}
\usepackage{amsmath}
\usepackage{amssymb}
\usepackage{epsfig}

\newcommand{\eqb}{\begin{equation}}
\newcommand{\eqe}{\end{equation}}
\newcommand{\bqn}{\begin{equation}}
\newcommand{\eqn}{\end{equation}}

\newcommand{\rd}{\mathrm{d}}
\newcommand{\bx}{{\bf x}}

\newcommand{\vare}{\varepsilon }
\newcommand{\pd}{\partial}

\begin{document}
\title{Mesoscopic quantum switching of a Bose-Einstein condensate in an optical
lattice governed by the parity of the number of atoms}
\author{ V. S. Shchesnovich }
\affiliation{ Centro de Ci\^encias Naturais e Humanas, Universidade Federal do ABC,
Santo Andr\'e,  SP, 09210-170 Brazil}

\begin{abstract}

It is shown that for  a $N$-boson system the parity of $N$ can be responsible for a
qualitative difference in the system response to variation of a parameter.  The
nonlinear boson model is considered, which describes tunneling of boson pairs
between two distinct modes $X_{1,2}$ of the same energy and applies to a
Bose-Einstein condensate in an optical lattice. By varying the lattice depth one
induces the parity-dependent quantum switching, i.e. $X_1\to X_2$ for even $N$ and
$X_1\to X_1$ for odd $N$, for arbitrarily large $N$. A simple scheme is proposed
for observation of the parity effect on the \textit{mesoscopic scale} by using the
bounce switching regime, which is insensitive to the initial state preparation (as
long as only one of the two $X_l$ modes is significantly populated), stable under
small perturbations  and requires an experimentally accessible coherence time.

\end{abstract}
 \maketitle

Mesoscopic quantum phenomena lie in between the big and small: the macroscopic
classical world and the microscopic quantum world. The  Bose-Einstein condensate
(BEC) is such a mesoscopic effect, i.e. a big ``matter-wave''. An order parameter
governed  by the Gross-Pitaevskii equation \cite{Legg,PS,MF} is usually attributed
to BEC. The order parameter corresponds to the mean-field theory, i.e. to the limit
of large number of bosons: $N\to\infty$ at a constant density. The latter, on the
other hand, is equivalent to the classical limit of the discrete WKB approach
\cite{Braun}, with $1/N$ playing the role of the Planck constant (see, for
instance, Refs. \cite{SK1,SK2}).

The mean-field limit  is a singular limit of the full quantum description and
suffers from deficiency, e.g. at a dynamic instability \cite{MF}, due to the
back-reaction of the quantum fluctuations \cite{AV}, or the appearance of the
Schr\"odinger cat-like states \cite{WT}.  In this connection one can mention the
``even-odd" effect, first predicted for the spin systems \cite{Spar} and observed
in the small ($S\sim 10$) magnetic molecular clusters as the parity-dependent
tunneling splitting \cite{SparExp}. The parity effect was also found in the decay
of the Josephson $\pi$-states \cite{Jpi} and in the boson-Josephson model
\cite{BEC2w}.  The tunneling splitting, however, decreases exponentially in $N$,
for $N\gg1$, restricting its observation to the sub-mesoscopic scale. One may
wonder whether it is possible to magnify the microscopic parity difference to a
mesoscopic scale and how? Such an effect would be also an interesting manifestation
of the singularity in the $N\to \infty$ limit of the discrete WKB.

The aim of this  rapid communication is to present a solution:  one must look
for a dynamic parity effect which allows for a  massive constructive quantum
interference. Moreover, the  feasibility  of the  experimental observation is
shown. The mesoscopic parity effect appears in the response to  variation of a
parameter in the nonlinear two-mode boson model \cite{SK1}, a nonlinear variant of
the celebrated boson-Josephson model \cite{bJ,SKPRL}.

The nonlinear two-mode boson model  is formulated as follows. Suppose that a
single-particle Hamiltonian $H_0$ has two equal energy states $X_1$ and $X_2$ and
that the interaction term
$H_\mathrm{int}=\frac{g}{2}\int\rd^3\bx\psi^\dag(\bx)\psi^\dag(\bx)\psi(\bx)\psi(\bx)$
in the many-body Hamiltonian $H = \int\rd^3\bx\psi^\dag(\bx)H_0\psi(\bx)
+H_\mathrm{int}$ is smaller than the energy gap of $H_0$ isolating the resonant
subspace. Projecting on the resonant  states, $\psi(\bx) = b_1\varphi_{X_1}(\bx) +
b_2\varphi_{X_2}(\bx)$, one arrives at the Hamiltonian: $H_\mathrm{int} =
\frac{g}{2}\left\{
\sum\chi_{i_1,i_2,j_1,j_2}b^\dag_{i_1}b^\dag_{i_2}b^{}_{j_1}b^{}_{j_2}\right\}$,
${i_1,i_2,j_1,j_2}\in\{1,2\}$, where
$\chi_{i_1,i_2,j_1,j_2}\equiv\int\rd^3\bx\,\varphi^*_{X_{i_1}}\varphi^*_{X_{i_2}}\varphi_{X_{j_1}}\varphi_{X_{j_2}}$.
We consider the situation when the bosons  hop between the modes $X_l$ by pairs,
i.e. when $\chi_{12jj}=0$ for  $j=1,2$. This type of coupling describes the
intraband tunneling of BEC in a square optical lattice \cite{SK1}, where the
resonant modes are the high-symmetry points of the Brillouin zone, $X_1 = (k_B,0)$
and $X_2 = (0,k_B)$, and the quasimomentum conservation makes $\chi_{12jj}$ vanish.
Moreover, it also applies to BEC on a rotated ring lattice \cite{RL}. The intraband
BEC tunneling Hamiltonian reads
\eqb \hat{H} = \frac{1}{2N^2}\left\{n_1^2+n_2^2
+\Lambda\left[4n_1n_2+(b_1^\dag b_2)^2 + (b_2^\dag b_1)^2 \right] \right\},
\label{EQ1}
\eqe
where  $n_j=b^\dag_jb_j$ and $\Lambda= \chi_{1122}/\chi_{1111}$ ($0\leq \Lambda\leq
1$)  is the only parameter  in the model (see for details Ref. \cite{SK1,SKPRL}).
The corresponding Schr\"odinger equation is cast as \mbox{$ih\pd_\tau |\Psi\rangle
= \hat{H}|\Psi\rangle$}, where  $h = 2/N$ is the effective Planck constant  and the
dimensionless  time $\tau = (2gN\chi_{1111}/\hbar)t$, with $g = 4\pi\hbar^2 a_s/m$,
depends on $N$ through the density only.

\begin{figure}[htb]
\begin{center}
\epsfig{file=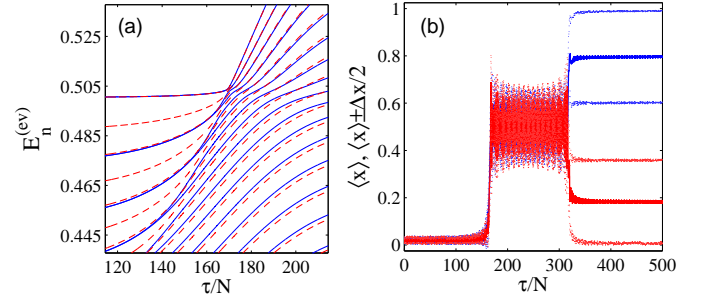,width=0.5\textwidth} \caption{(Color online) Panel (a): the
energy level structure of $H^{(\mathrm{ev})}$ for $N=40$ (solid lines) and $41$
(dashed lines). Panel (b): the average ratio of the $X_1$-mode occupation number
$\langle x\rangle =\langle b_1{\!\!}^\dag b_1\rangle/N$ (dark solid lines) and
$\langle x\rangle -\frac{\Delta x}{2}$, $\langle x\rangle+\frac{\Delta x}{2}$
(light dashed lines), with  \mbox{$\Delta x = \langle(b_1{\!\!}^\dag b_1/N-\langle
x\rangle)^2\rangle^{1/2}$}. The upper lines (solid and dashed) correspond to $N=40$
and the lower ones to $N=41$. The initial  state is a Gaussian \mbox{$|\Psi\rangle
= A_\sigma\sum_{k=0}^N e^{-k^2/\sigma^2}|k,N-k\rangle$}, where $\sigma = 4$. Here
$\Lambda(\tau)=\Lambda_1 +
1/2(\Lambda_2-\Lambda_1)\left[\tanh\left(\kappa[\tau-\tau_1] \right) - \tanh\left(
\kappa[\tau-\tau_2] \right)\right]$ with  $\kappa = 10^{-4}$, $\tau_1 = 7000$ and
$\tau_2 = 12800$.}
\label{FG1}
\end{center}
\end{figure}

Hamiltonian (\ref{EQ1}) features \cite{SKPRL} a quantum phase transition at the top
of the spectrum  related to the mean-field symmetry-breaking bifurcation between
the stationary point $(\langle b^\dag_1b_1\rangle/N = 1/2,\phi\equiv \arg\langle
(b_2^\dag)^2 b_1^2\rangle=0)$, corresponding to the equally populated $X_{l}$
modes, which is stable for $\Lambda
>\Lambda_c=1/3$, and  the selftrapping  stationary points ($\langle b^\dag_1b_1\rangle/N \ll1
$ or $\langle b^\dag_2b_2\rangle/N \ll1 $, $\phi$ undefined), stable for
$\Lambda<\Lambda_c$. It also has a parity-dependent energy spectrum (see also Fig
\ref{FG1}(a)). There are two invariant subspaces corresponding  to the even and odd
occupation numbers $k$ in the Fock basis, i.e. $|\Psi\rangle =
\sum_{k=0}^NC_k|k,N-k\rangle$, with $|k,N-k\rangle\equiv
\frac{(b^\dag_1)^k(b^\dag_2)^{N-k}}{\sqrt{k!(N-k)!}}|\mathrm{vac}\rangle$. The
projections of $\hat{H}$ on  the even ($\mathrm{s}=0$ or ``ev") and odd
($\mathrm{s}=1$ or ``od") subspace are given as
\eqb
H^{(\mathrm{s})} = \begin{pmatrix}{\alpha}_s & {\beta}_{s+1} & 0 & \ldots & 0\\
{\beta}_{s+1} & {\alpha}_{s+2} & \ddots & \ddots & \vdots\\
0 & \ddots & \ddots & \ddots & 0 \\
\vdots & \ddots & \ddots & \ddots & {\beta}_{2L-1-s}\\
0 & \ldots & 0 & {\beta}_{2L-1-s} & {\alpha}_{2L-s} \end{pmatrix},
\label{EQ2}\eqe
where, respectively for even and odd $N$: $L = N/2$ and $L = (N-1)/2$ in the case
of $H^{(\mathrm{ev})}$, while $L = N/2$ and $L = (N+1)/2$ in the case of
$H^{(\mathrm{od})}$. Here ${\alpha}_k =
(2\Lambda-1)\frac{k}{N}\left(1-\frac{k}{N}\right)$ and \mbox{${\beta}_k =
\frac\Lambda2\left[\frac{k}{N}\left(\frac{k}{N}+\frac1N\right)
\left(1-\frac{k}{N}\right)\left(1-\frac{k}{N}+\frac1N\right)\right]^{1/2}$.}

When $N$ is odd we have  $PH^{(\mathrm{ev})}P = H^{(\mathrm{od})}$, where  $P =
\mathrm{diag}(1,\ldots,1)^T$ (i.e. the transposed $\openone$). In this case
$P|E^{(\mathrm{ev})}_j\rangle = |E^{(\mathrm{od})}_j\rangle$, i.e. the energy
levels are doubly degenerate. On the other hand, this is  a consequence of the
Kramers theorem \cite{KTh}. Indeed, Hamiltonian (\ref{EQ1}) is equivalent to a spin
model \mbox{$H_S = (1-\Lambda)S^2_z+2\Lambda S^2_x$} with the total spin $S=N/2$,
if we associate $S_x = (b^\dag_1b_2+b^\dag_2b_1)/2$, $S_y =
(b^\dag_1b_2-b^\dag_2b_1)/2i$ and $S_z = (b^\dag_1b_1-b^\dag_2b_2)/2$.

When $N$ is even  the projected Hamiltonian $H^{(\mathrm{s})}$ is invariant under
the exchange symmetry, i.e. $PH^{(\mathrm{s})}P = H^{(\mathrm{s})}$, $\mathrm{s}
\in\{\mathrm{0,\; 1\}}$. One control parameter can not cause the  energy level
degeneracy \cite{NW}, thus the eigenvectors of $H^{(\mathrm{s})}$ must satisfy the
exchange symmetry, namely $P|E^{(\mathrm{s})}_j\rangle =
(-1)^{j+\frac{N}{2}+1-s}|E^{(\mathrm{s})}_j\rangle$, $j = 1,\ldots,
\frac{N}{2}+1-s$. For a finite $\Lambda_c-\Lambda>0$ the eigenvalues of
$H^{(\mathrm{s})}$ appear in the form of   very narrow doublets (see also Fig.
\ref{FG1}(a)) due to the ``selftrapping states'' being strongly localized at the
respective $X_l$ mode (e.g. typically $|\langle k,N-k|\Psi\rangle|^2\le 10^{-6}$
for  $k>25$ \cite{SKPRL}). Consequently, each $H^{(\mathrm{s})}$ also has the
quasi-degenerate spectra (since $H^{(\mathrm{ev})} - H^{(\mathrm{od})} \propto
\frac1N$) which become finer for larger deviations $\Lambda_c-\Lambda$, exactly as
the numerics indicates.

Consider now the following experimentally realizable  setup:  initially just one of
the $X_l$ modes is significantly populated (to achieve this one can use the
non-adiabatic loading \cite{Blochload} into one of the two resonant Bloch states of
the lattice with $\Lambda_1<1/3$). By varying the lattice parameter (e.g. by changing
the lattice depth) between $\Lambda_1$ and $\Lambda_2$, with $\Lambda_2>\Lambda_c$,
one drives the system across the phase transition and back to force a
switching-like dynamics between the selftrapping states at the $X_{l}$ modes.
Remarkably, for a general initial state 
localized at just one $X_l$ mode, e.g. $|\Psi\rangle = \sum_{k=0}^{K\ll
N}C_k|k,N-k\rangle$, where the distribution $C_k$ is not important,   the result qualitatively depends on the parity of $N$, see Fig.~\ref{FG1}(b). Note that the switching is between the Bloch modes
with orthogonal Bloch vectors: $X_1=(k_B,0)$ and $X_2=(0,k_B)$.

\begin{figure}[htb]
\begin{center}
\epsfig{file=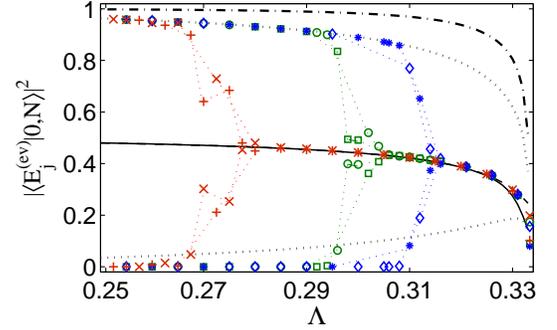,width=0.4\textwidth} \caption{(Color online) The probabilities
$|\langle E^{(\mathrm{ev})}_{j}|0,N\rangle|^2$, $j=L+1$ and $j=L$, given for $N=40$
by the solid and dashed lines (almost coincide), for $N=60$ by the ``$+$'' and
``$\times$'', for $N=80$ by the ``{\tiny $\square$}'' and ``$\circ$'' and for
$N=100$ by the ``$* $'' and ``$\diamond$'' symbols. The upper dash-dotted line
gives $\sum_{m=0}^{3}|\langle E^{(\mathrm{ev})}_{L+1-m}|0,N\rangle|^2$. The upper
and lower dotted lines give, respectively,  $|\langle
E^{(\mathrm{ev})}_{L+1}|0,N\rangle|^2$ and $\sum_{m=1}^{3}|\langle
E^{(\mathrm{ev})}_{L+1-m}|0,N\rangle|^2$ for odd $N$.  The data, connected by the
dotted lines to guide the eye, that lie off the central curve are due to the
quasi-degeneracy on the order of round-off error.}
\label{FG2}
\end{center}
\end{figure}

In the adiabatic limit the mechanism of the switching for a localized initial
distribution $C_k$ can be understood by considering separately the even and odd
invariant subspaces. For simplicity, consider the initial state $|\Psi\rangle =
|0,N\rangle \in H^{(\mathrm{ev})}$ (the case of $|\Psi\rangle=|1,N-1\rangle \in
H^{(\mathrm{od})}$ is similar). Fig.~\ref{FG2} shows that  there are but few
significant terms (from the top of the spectrum) in the expansion of the initial
state over the eigenvectors (recall the strong localization of the eigenvectors at
$k=0$ and $k=N$ for $\Lambda<\Lambda_c$). For even $N$ there are the quantum beats,
i.e. oscillations of the populations, between the successive pairs of eigenstates
from the top of the energy spectrum. This can be understood as  tunneling between
the $X_{l}$ modes. The switching occurs for the odd-$\pi$ phase differences between
the first few pairs of the successive top energy eigenstates, i.e.
$\Delta\theta_{L+1-2j}\equiv \frac{1}{h}\int\limits_0^\tau\rd t\,
[E^{(\mathrm{s})}_{L+1-2j}(t)-E^{(\mathrm{s})}_{L-2j}(t)] \approx (2m_j-1)\pi$,
$j=0,1,...\,$. For odd $N$, on the other hand, the eigenvectors break the
$P$-symmetry and only those localized at $X_1$ acquire nonzero amplitudes (i.e. no
tunneling). Thus, a high visibility parity effect requires at least few top dynamic
phase differences in both the even and odd subspaces  (for even $N$) to be odd in
$\pi$, which,  in the general case, can not be satisfied  by adjusting
$\tau_2-\tau_1$ in Fig.~\ref{FG1}.

The Landau-Zener-Majorana (LZM) transitions  between the instantaneous energy
levels occur for the non-adiabatic variation of $\Lambda$. In the instantaneous
eigenvector basis, $|\Psi^{(\mathrm{s})}(\tau)\rangle =
\sum_{j}c_j(\tau)\exp\left[-\frac{i}{h}\int\limits_0^\tau\rd
t\,E^{(\mathrm{s})}_j(t)\right]|E^{(\mathrm{s})}_j(\tau)\rangle$, we have
\eqb
\frac{\rd c_j}{\rd \tau} = \frac{\rd\Lambda}{\rd\tau}\sum_{l\ne j}c_l\frac{\langle
E^{(\mathrm{s})}_j |\frac{\rd \hat{H}}{\rd
\Lambda}|E^{(\mathrm{s})}_l\rangle}{E^{(\mathrm{s})}_j -E^{(\mathrm{s})}_l}e^{-\frac{i} {h}\int\limits_0^\tau\rd
t\,[E^{(\mathrm{s})}_j-E^{(\mathrm{s})}_l]},
\label{EQ3}\eqe
where it was used that  $\langle E^{(\mathrm{s})}_j|\frac{\rd }{\rd
\tau}|E^{(\mathrm{s})}_j\rangle = 0$ ($H^{(\mathrm{s})}$ does not have energy
degeneracy). Due to the  exchange symmetry $P$, in the even $N$ case the LZM
tunneling occurs only between the levels $E^{(\mathrm{s})}_j$ with the same parity
of $j$, whereas for odd $N$ there is a small coupling also between the adjacent
levels, since $H^{(\mathrm{s})}_{2L+1} - H^{(\mathrm{s})}_{2L} \propto \frac1N$.
The LZM result \cite{LZM} states that $|c_l|^2 \sim \exp\left(-\frac{\pi(\Delta
E)^2}{h\left(\frac{\rd \Delta E}{\rd \tau}\right)}\right)$, for $l\ne j$. Hence,
the lower limit on the adiabatic  time scale, i.e. $|c_l|^2\ll 1$, can be
determined from the difference between the top energy levels
$E^{(\mathrm{s})}_{n+1} -E^{(\mathrm{s})}_n \propto \frac{1}{N^2}$ for $\Lambda \to
\Lambda_c$ \cite{SKPRL} (see also below). We get $\tau_\mathrm{ad}\gg N$ (the time
scale for a finite phase difference $\Delta\theta_j$ is $\tau_\mathrm{ph} \sim N$,
since $h=2/N$). Therefore,  the adiabatic case also requires an extremely long
coherence time and thus it is not realistic at all.

\begin{figure}[htb]
\begin{center}
\epsfig{file=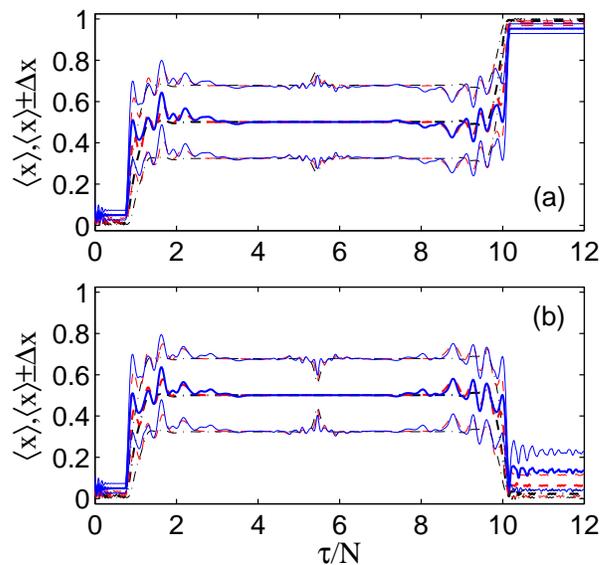,width=0.45\textwidth} \caption{(Color online) The bounce
switching for the initial state $|\Psi\rangle =
A_\sigma\sum_{k=0}^Ne^{-k^2/\sigma^2}|k,N-k\rangle$. Panel (a) $N=200$ and panel
(b) $N=201$. In both panels, the average ratio $\langle x\rangle$ (the thick lines)
with the side lines  $\langle x\rangle -\frac{\Delta x}{2}$ and $\langle
x\rangle+\frac{\Delta x}{2}$ (the thin lines) are given for $\sigma = 0.05N$
(dash-dotted lines), $\sigma = 0.1N$ (dashed lines), and $\sigma= 0.2N$ (solid
lines). Here $\Lambda(\tau)$ is as in Fig.~\ref{FG1} with $\kappa=8$, $\Lambda_1 =
0.25$, $\Lambda_2 = 1/3$, and $\tau_2-\tau_1 = 3N\pi$. }
\label{FG3}
\end{center}
\end{figure}

In the other limit, Fig. \ref{FG3}, when  $\Lambda(\tau)$ is a step-like function
(e.g. similar to the one used in Fig. \ref{FG1}(b), but with $\kappa\sim 1$ or
larger) taking two values $\Lambda_1<\Lambda_c$ and $\Lambda_2=\Lambda_c$, the
switching regime, the \textit{bounce switching} for below, has all the needed
properties.

Indeed, the bounce switching possesses  the $N$-scaling property, thus it \textit{survives}
in the $N\to\infty$ limit, and  is  insensitive  to the initial distribution
$C_k$ (localized at just one $X$-mode). These features originate from the fact that all the dynamic phases are odd in $\pi$. Indeed,  Hamiltonian (\ref{EQ1}) in the
coherent basis $a_{1,2} = (b_1 \mp ib_2)/\sqrt{2}$ reads $ \hat{H}=
\frac{2\Lambda}{N^2} a^\dag_1a_1a^\dag_2 a_2 + \frac{1-3\Lambda}{4N^2}
\left(a_1^\dag a_2 +a_2^\dag a_1\right)^2 + \mathrm{const}$ with the energies
$E_{\ell}(\Lambda_c) = \frac{2}{3N^2}(\ell-1)(N-\ell+1)$,
$\ell=1,...,\left[\frac{N}{2}\right]+1$. Hence, the dynamic phase differences are
given as $\Delta\theta^{(\mathrm{s})}_{L+1-s-2n} = \frac{4n+2s+1}{3N}\Delta\tau_c$,
$n=0,1,2,\ldots$, where $\Delta\tau_c$ is the hold time at $\Lambda_c$. By setting
$\Delta\tau_c = 3\pi N$ one obtains $\Delta\theta^{(\mathrm{s})}_{L+1-s-2n}
=(4n+2s+1)\pi$.

The  actual physical time, $t = t_\mathrm{ph}\frac{\tau}{N}$ with $t_\mathrm{ph}
\equiv\frac{\hbar}{2g\chi_{1111}}$, is  $N$-independent due to the scaling property
$\Delta\tau_c = 3\pi N$. For a condensate in a square lattice of $n$ cites of size
$d$ in the tight transverse trap with the oscillator length $a_\perp$ the coherence
time is $\Delta t \sim \frac{m d^2 a_\perp n}{\hbar a_s}$. For ${}^{87}$Rb, $n=
64$, $d=0.5\mu$m and $a_\perp=0.1\mu$m the  required coherence time is $\Delta
t\sim 0.3\,\mathrm{s}$.  The switching  can be detected by releasing the optical
lattice and observing the direction of the interference pattern. For the lattice $V
= V_0\left[\cos(2k_Bx)+\cos(2k_By)\right]$ the parameter range is $V_0 = 0.58E_R$ for
$\Lambda = 0.25$ and $V_0 = 0.3573E_R$ for $\Lambda = 1/3$. Finally, the
applicability condition for the two-mode model  $E_\mathrm{NL}/E_R\ll1$, where $E_R
\equiv \hbar^2k_B^2/2m$ and $E_\mathrm{NL} \equiv gN\chi_{1111}$, can be cast as $
N\ll na_\perp/a_s$.

Experimental observation requires stability of the dynamic parity effect under
perturbations. To have an idea  on  the bounce switching stability, consider first
the general perturbation within the two-mode model:
\eqb
\hat{H}_\mathrm{pert} = \frac{\vare}{N}(n_1-n_2) +
\frac{J}{N}(b^\dag_1b_2+b^\dag_2b_1),
\label{EQ5}\eqe
where $\vare$ and $J$ (given in terms of the nonlinear energy $E_\mathrm{NL}$)
account for the imperfections of the optical lattice and for a magnetic trap (but, see
below). The nonlinear interaction terms $g\chi_{12jj}(b^\dag_1b^\dag_2[b^2_1
+b^2_2] + h.c.)$ discarded when deriving Eq. (\ref{EQ1}) also reduce to the
$J$-part in Eq. (\ref{EQ5})  with $J = \chi_{12jj}/\chi_{1111}$. The first term in
Eq. (\ref{EQ5}) preserves the decomposition of the model as in Eq. (\ref{EQ2}),
whereas the second one breaks it and  both break the exchange symmetry $P$. In the
$a$-operator basis $\hat{H}_\mathrm{pert} =
\frac{\vare}{N}(a^\dag_1a_2+a^\dag_2a_1) -\frac{iJ}{N}(a^\dag_1a_2-a^\dag_2a_1)$
and  the matrix elements of $\hat{H}_\mathrm{pert}$ between the eigenstates of
$\hat{H}$ are much smaller than the energy  differences  at $\Lambda_c$ if
$|\vare-iJ|\ll 1/N^2$. However, extensive numerical simulations show that for $N\le
(\vare^2+J^2)^{-1/4}$ the system  still exhibits essentially the same parity effect
as in Fig.~\ref{FG3}. For $N \le |\Lambda_2-1/3|^{-1/2}$, the parity effect is also
insensitive to an imprecise tuning of $\Lambda_2$ in Fig.~\ref{FG3} to the critical
value.

An additional weak magnetic trap $V_\mathrm{tr} = m\omega^2(x^2+y^2)/2 +
V_\perp(z)$  is always a part of the experimental setup, leading to the $J$-term in
Eq. (\ref{EQ5}). However, this contribution is exponentially small. Indeed, for a
weak trap  $\ell^2\equiv\hbar/m\omega\sim nd^2\gg d^2$  the unperturbed Bloch wave
$\varphi_{\mathbf{k}}(\bx)$ acquires a factor given by a product of a polynomial
and a Gaussian in $(x,y)$, i.e. $\varphi^{(V_\mathrm{tr})}_{\mathbf{k}}(\bx) =
G(z)P(x,y)\exp\{-\frac{x^2+y^2}{2\ell^2}\}\varphi_{\mathbf{k}}(\bx)$. The Gaussian
factor defines the order of  the non-diagonal matrix element of $V_\mathrm{tr}$
between the two resonant Bloch waves. We have
\eqb
\int\rd^3\bx\,\varphi^{(V_\mathrm{tr})}_{(k_B,0)}V_\mathrm{tr}\varphi^{(V_\mathrm{tr})}_{(0,k_B)}
\sim \hbar\omega\exp\left\{-\frac{\pi^2\ell^2}{4d^2}\right\},
\eqe
therefore $J$ in Eq.~(\ref{EQ5}) reads $J \sim
(\hbar\omega/E_\mathrm{NL})\exp\{-\pi^2n/4\}$. To bound the pre-exponential factor
$\hbar\omega/E_\mathrm{NL} \sim nd^2a_\perp/(\ell^2a_sN)\sim a_\perp/(a_sN)$ one
needs $N\sim a_\perp/a_s$ compatible with the applicability condition $N\ll
na_\perp/a_s$.

There still remains to consider the transitions to the non-resonant modes, due to
the nonlinear term of the full boson Hamiltonian. The latter, however, preserve the
quasi-momentum (with the exponentially small correction due to the magnetic trap)
and, hence, the parity of $N$, since the bosons leave the resonant modes by pairs.
More detrimental than the setup imperfections considered above is the loss of
atoms, for instance the scattering of  BEC atoms with the cloud of hot atoms, which
will wash out the parity effect. To prevent this, a smaller then in a usual BEC
number of cold atoms can be used, e.g. $N$ on the order of few hundred atoms.

In conclusion, for a flexible control parameter, the  nonlinear two-mode boson
model possesses a bounce switching regime with the \textit{qualitatively} different
outcome of switching for even and odd $N$. This regime is insensitive to the
initial state preparation (with just one resonant mode being significantly
populated), shows  stability to small perturbations and requires an experimentally
accessible coherence time, thus allowing for observation of the even-odd effect on
the mesoscopic scale. As a general perspective, one can observe  that the nonlinear
two-mode boson  model is a  nonlinear variant of the two-site Bose-Hubbard
Hamiltonian and that the second-order  tunneling applies also to the dynamics of
the repulsively bound atom pairs in an optical lattice \cite{AtPairs,SecordTunl}
and to the case of strong interactions reaching the fermionization limit \cite{Th}.

\acknowledgments   This work was supported by  the FAPESP  and CNPq of Brazil.

\end{document}